\documentclass{osa-article}

\journal{osajournal}

\articletype{Research Article}

\usepackage{lineno}

\usepackage{lipsum}
\usepackage{graphicx}

\usepackage{todonotes}
\usepackage{siunitx}
\usepackage{amsmath}
\usepackage{comment}
\usepackage[utf8]{inputenc}
\usepackage[T1]{fontenc}
\usepackage{newtxtext,newtxmath}

\begin{document}

\title{Discretization of Annular-Ring Diffraction Pattern for Large-Scale Photonics Beamforming}

\author{Aroutin Khachaturian\authormark{*}, Reza Fatemi, Artsroun Darbinian, and Ali Hajimiri}

\address{California Institute of Technology, USA\\}
\email{\authormark{*}akhachat@caltech.edu} 



\begin{abstract}
A solid-state active beamformer based on the annular-ring diffraction pattern is proposed for an integrated photonic platform. Such a circularly symmetric annular-ring aperture achieves radiating element limited FOV. Furthermore, it is demonstrated that a multi-annular-ring aperture with a fixed linear density of elements maintains the beam efficiency for larger apertures while reducing the beamwidth and side-lobe-level (SLL). A $255$-element multi-annular-ring OPA with active beamforming is implemented in a standard photonics process. $510$ phase and amplitude modulators enable beamforming and beam steering using this aperture. A row-column drive methodology reduces the required electrical drivers by more than a factor of 5. 
\end{abstract}


\section{Introduction}

Integrated solid-state photonic beamformers (optical phased arrays or OPAs) have the potential to reduce the cost, size, and implementation complexity of many photonic systems compared to their bulk optics and MEMS counterparts \cite{TUANTRANONT:01, Wang:19}. These solid-state beamformers have been recently demonstrated for LiDAR \cite{Aflatouni_NI:15, Hashemi:2021}, photonic beam steering \cite{Aflatouni:15, Fatemi:17, Ashtiani:19}, medical imaging \cite{Eggleston:18}, and remote sensing \cite{White2020} applications. In particular, standard silicon photonics processes can further reduce the cost and increase the yield and reliability of such systems \cite{Fatemi:17, Poulton:19, Miller:18, Watts:Nature}. However, there are several challenges associated with the realization of large-scale solid-state beamformers. 

Most integrated photonic dielectric waveguides and radiators have a minimum size and spacing on the order of the wavelength. This leads to minimum pitch and spacing constraints that lead to reduced field-of-view (FOV) and increased grating lobes as the size of the array aperture increases. There are two categories of architectures that have been utilized to overcome this problem. One class of architectures uses a wavelength-sensitive 1D-grid array of radiating elements to steer the beam in one direction and sweep the wavelength of the laser to steer the beam in the perpendicular direction \cite{Poulton:19, Miller:18, Abediasl:2017, Dostart:20, Poulton:17}. This method removes the planar routing restriction at the cost of increased laser source and system complexity. Typically, a broadly tunable laser source (around \SI{100}{\nano \meter} of wavelength tunability) is required to achieve a moderate FOV (around $20^\circ$) \cite{Watts:512, Ma:20, Dostart:20}. Such tunable lasers are more complex and hence more costly compared to their single wavelength counterparts. Furthermore, such OPAs cannot offer wavefront control in the steering direction controlled by the wavelength. \par
 
Another class of architectures utilizes sparse array synthesis techniques to construct a 2D-grid array of radiating elements that permit routing the signals of the inner elements of the array \cite{Fatemi:17}. Compared to their equal-size aperture, half-wavelength spacing counterparts, sparse arrays reduce the number of the radiating elements and phase shifters required, thus reducing array control complexity, power consumption, heat dissipation, and system cost. Furthermore, such arrays can operate with a fixed-wavelength laser source. These 2D-grid OPAs are advantageous over their 1D-grid aperture counterparts since they can offer full wavefront control with a fixed-wavelength laser. However, sparse placement of the elements reduces the array gain and beam efficiency (also known as the "sparse array curse"). Furthermore, most sparse array synthesis techniques achieve target beam performance by the randomized placement of elements in a 2D grid given the constraints on the number of elements and planar signal routing. For example, genetic algorithms have been used for sparse array synthesis \cite{Fatemi:17,GeneticSparse:Oneill,GeneticSparse:Haupt}.
This optimization process is computationally expensive for larger aperture sizes with an increased number of elements and there is no guarantee of arriving at the optimal design.

\begin{figure}[!t]
	\includegraphics[width=1\linewidth]{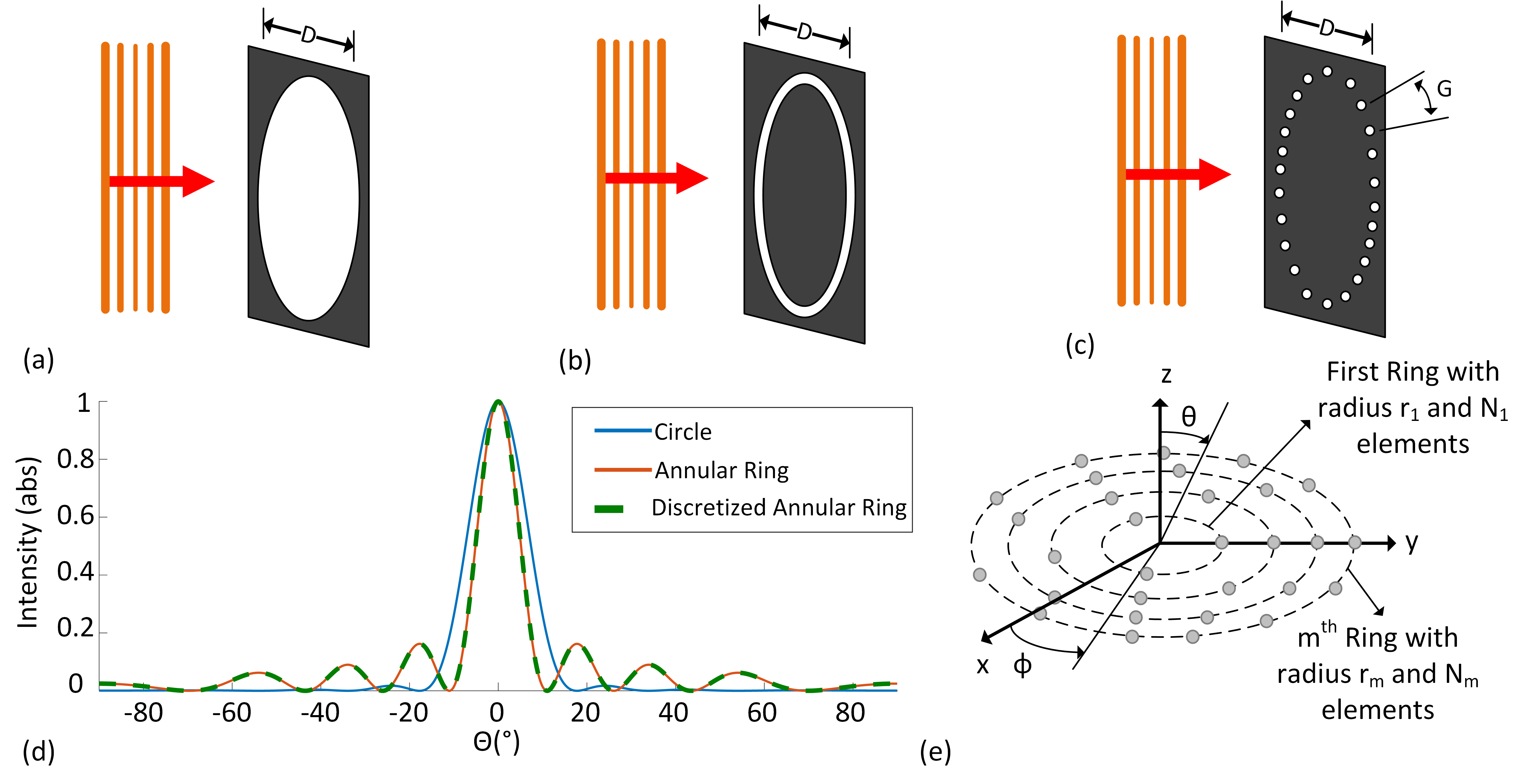}
	\caption{Diffraction pattern of circularly symmetric apertures with diameter $D=2\lambda$ when illuminated by a plane wave. (a) Circular aperture. (b) An annular-ring aperture. (c) Discretization of the annular-ring with $25$ points ($G\approx \lambda/2$ is the arc-length distance between two elements) (d) Diffraction pattern cross-section of apertures (a),(b), and (c). (e) Generalization of discretized multi-annular-ring apertures equivalent to circular-grid phased arrays.}
	\label{RingSlifDiff}
\end{figure}

One method to reduce the complexity of finding optimum sparse apertures is to take advantage of the basic properties of circularly symmetric apertures. The ideal symmetric aperture is circular, as shown in Fig. \ref{RingSlifDiff}(a). The beam pattern of such an aperture can be computed by taking the Fourier transform of a circle:
\begin{equation}
|P_{circ}(\theta)|= |\mathcal{F}\left[\mathrm{circ}(D) \right]| = \frac{J_1\left(\frac{kD}{2} \sin(\theta)\right)}{\frac{kD}{2}\sin(\theta)}
\label{AF_Ring}
\end{equation} 
where $P$ is the radiation pattern, $\theta$ is the elevation angle, $D$ is the diameter of the circle, $k$ is the wave number and $J_1$ is the first order Bessel function of the first kind.
The beam intensity cross-section of such aperture is shown in Fig. \ref{RingSlifDiff}(d), as is described by function $J(x)/x$. Another circularly symmetric aperture is the annular-ring aperture (Fig. \ref{RingSlifDiff}(b)), which can be modeled by blocking the center region of a circular aperture. The far-field beam pattern for such an aperture is given by the Fourier transform of difference of two circular apertures of diameter $D$ and $D-\delta$, where $\delta$ is small:
\begin{equation}
|P_{Ring}(\theta)|= |\mathcal{F}\left[\mathrm{circ}(D) - \mathrm{circ}(D-\delta) \right]|= J_0\left(\frac{kD}{2} \sin(\theta)\right)
\label{AF_AnnularRing}
\end{equation} 
where $J_0$ is the zeroth order Bessel function of first kind.
The beam intensity of this cross-section is shown in Fig. \ref{RingSlifDiff}(d). 

These beam patterns show that annular-ring apertures reduce the beamwidth at the cost of increased side-lobe level (SLL) compared to their circular counterparts that have an aperture with a significantly larger area. More importantly, annular-ring apertures do not produce grating lobes. This annular-ring aperture can be realized in a planar photonics process. We can physically implement an OPA capable of active beam steering by placing radiating elements on this annular ring at half-wavelength spacing \ref{RingSlifDiff}(c). This discretized annular-ring array pattern approximates to the pattern of a continuous annular-ring aperture \ref{RingSlifDiff}(d) for sufficiently large number of elements. It is possible to combine multiple such discretized annular rings to improve OPA beamwidth and SLL, as shown in Fig. \ref{RingSlifDiff}(e). In phased array theory, such multi-annular-ring apertures can be categorized under circular-aperture arrays  \cite{Balanius:2015,CPA:1999,Zhang_Pen:18,UCA:2011,khodier2009linear,austeng2002sparse}. Such symmetric apertures are advantageous over their rectangular counterparts since they exhibit minimal disturbance to the beamwidth and SLL when scanned azimuthally over the entire plane \cite{UCABalanis:2005}. 

In this paper, we present the first demonstration of such OPAs in a silicon photonics platform. In the next section, we analyze the beamforming characteristics of such annular rings in the context of planar photonics platforms. Afterwards, we present a 255-element silicon photonics implementation of such an annular-ring OPA with full amplitude and phase control for individual radiators. We simplify the electrical drive interconnect complexity of such an OPA by using a row-column drive which reduces the total number of electrical interconnects from 510 nodes to 100 nodes. Finally, we discuss the beamforming optimization methodology of this large-scale OPA and demonstrate the beamforming and beam steering capability of this OPA.

\section{Analysis of Annular-Ring Apertures}      

The discretized multi-annular-ring aperture can be generalized by creating $M$ concentric rings, each with $N_m$ isotropic radiators equidistant in arc length, where $m \in {1,2,...M}$ (Fig. \ref{RingSlifDiff}(e)). Using the expression for the far-field radiation pattern or array factor of a single annular-ring aperture from \cite{Balanius:2015}, we can write the generalized array factor for a multi-annular-ring aperture as
\begin{equation}
|P(\theta,\phi,\theta_0,\phi_0)|= \sum_{m=1}^{M} \sum_{n=1}^{N_m} I_{m,n} e^{jkr_m \left[\sin(\theta) \cos(\phi-\phi_{m,n})-\sin(\theta_0) \cos(\phi_0-\phi_{m,n})\right]}
\label{Eq_Balanus}
\end{equation} 
where $I_{m,n}$ is the $E$-field intensity for the particular element, $r_m$ is the radius of the $m$th ring, $\phi_{m,n}=2\pi n/N_m$ is the angular position of the $n$th element in the aperture, and $(\theta_0,\phi_0)$ pair describes the direction of the main beam. While amplitude apodization will help with SLL reduction, we limit this analysis to uniform excitation, $I_{m,n}=1$, in the broad-side direction, $(\theta_0,\phi_0)=(0,0)$. 

\begin{figure}[!b]
	\includegraphics[width=1\linewidth]{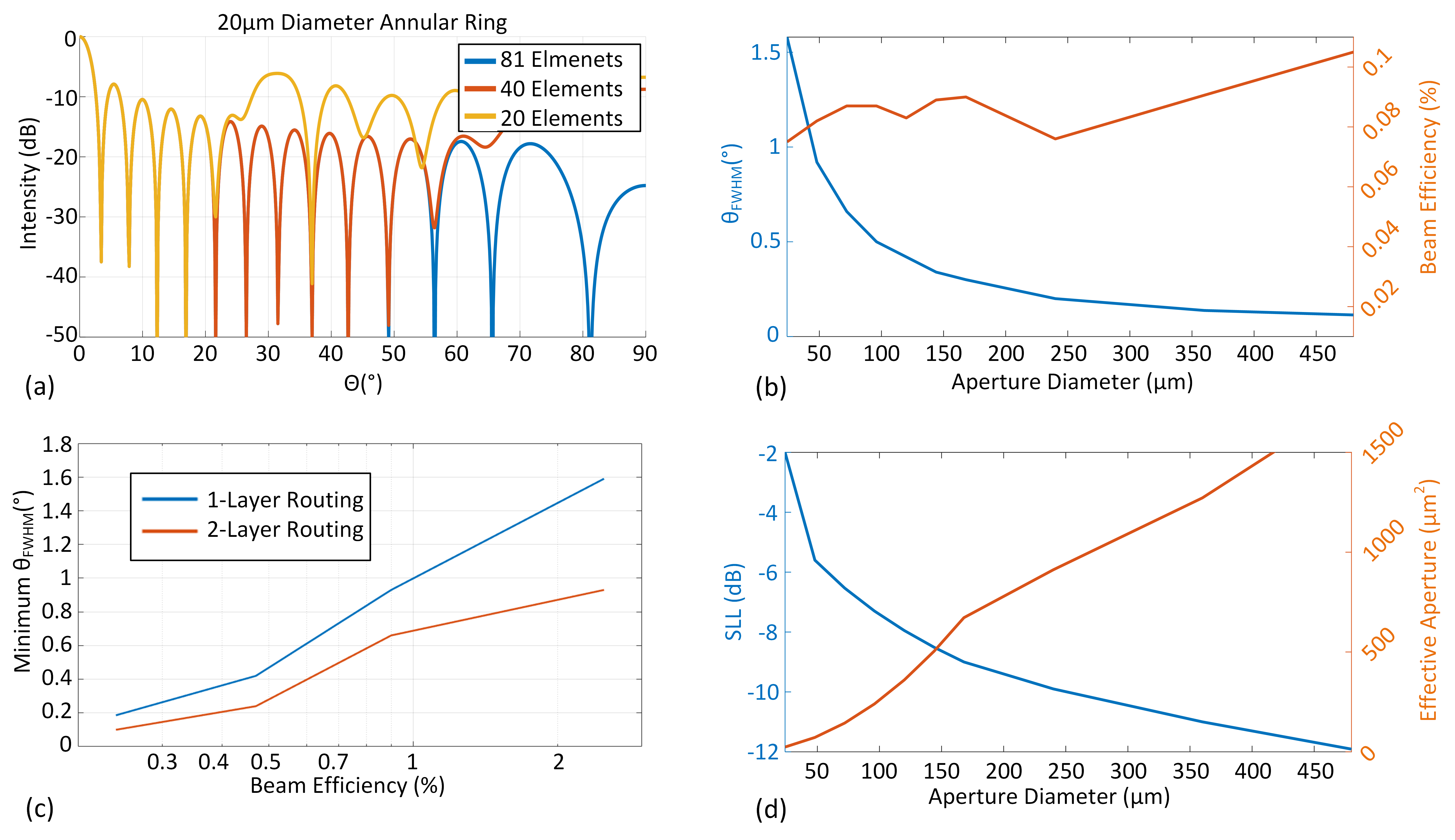}
	\caption{ (a) Effect of annular-ring discretization on the far-field array factor. A $20\mu m$ diameter ring is plotted for a continuous annular-ring-slit aperture (or half wavelengths-spacing elements), discretized with 40 isotropic radiators, and discretized with 20 isotropic radiators. (b) Beamwidth and \SI{3}{\decibel} beam efficiency trends as a function of phased array aperture diameter. (c) Minimum beamwidth as a function of \SI{3}{\decibel} beam efficiency for linear density multi-annular-ring OPAs at the planar routing limits for single-layer and two-layer photonics process. (d) SLL and effective aperture as a function of aperture diameter.}
	\label{Analysis1}
\end{figure}

\subsection{Approximation of Continuous Annular-Ring Aperture}
Using (\ref{Eq_Balanus}), we can compute the diffraction pattern of the continuous annular-ring aperture in Fig. \ref{RingSlifDiff}(b) by assuming $M=1$ and replacing summation of the discrete points by integration over the entire circumference. 
\begin{equation}
|P(\theta)|= \int_0^{2\pi} e^{jkr \sin(\theta) \cos(\phi_{n})}d\phi_{n} =2\pi J_0\left(k r \sin(\theta)\right) 
\label{Eq_SumGeneral_6}
\end{equation} 
After normalizing (\ref{Eq_SumGeneral_6}), the beam pattern is a Bessel function of order zero similar to the diffraction pattern calculation for an annular ring in (\ref{AF_AnnularRing}).
\par
As shown in Fig. \ref{RingSlifDiff}(d), placing discretized elements at distances up to half-wavelength spacing is equivalent to the continuous case with a Bessel function beam pattern. If the distance (over the arc) between the radiating elements increases to larger values, the Bessel function approximation holds for a narrower FOV. This is depicted in Fig. \ref{Analysis1}(a). Here, an annular-ring aperture with \SI{20}{\micro \meter} diameter and \SI{1.55}{\micro \meter} wavelength is plotted for continuous/half-wavelength case ($81$ elements), $40$, and $20$ radiating elements. For the case of $40$ radiating elements (\SI{1.6}{\micro \meter} arc length spacing), the Bessel approximation is valid up to $58^\circ$. For the case of $20$ radiating elements (\SI{3.1}{\micro \meter} arc length spacing), the Bessel approximation is valid for $\theta$ up to $23^\circ$. Furthermore, as the arc length spacing increase, the SLL increases.

\subsection{Annular-Ring Apertures with a Fixed Linear Density}
\label{LinearDensityTheory}
This multi-annular-ring aperture given by (\ref{Eq_Balanus}) is circularly symmetric, and as a result, the placement of the radiating elements can be optimized for desired performance parameters with reduced computational complexity compared to their rectangular grid sparse apertures. Such multi-annular-ring apertures can be analyzed for any of the beamforming parameters such as beamwidth, SLL, beam efficiency (\SI{3}{\decibel} beam power over total radiated power), and element count. In this work, we limit our analysis of annular-ring apertures to apertures with fixed linear density. In other words, we assume a constant elements per arc length density ($N_m=mN_1$) and linearly increase the ring radii ($r_m=mr_1$). 
\par
For such apertures with linear density, the planar photonic process parameters set the elements per arc length density as well as the minimum radius limits. For silicon photonics processes, the waveguides are typically \SI{500}{\nano \meter} wide with a minimum pitch of \SI{1}{\micro \meter} to reduce the electromagnetic coupling between the adjacent waveguides. Radiating elements can also be constructed at the size of the waveguides. For a fixed linear density, the design values $r_1$ and $N_1$ set the beam efficiency of the aperture. Assuming $r_1=\SI{40}{\micro \meter}$ and $N_1=20$, we can plot the beamwidth, beam efficiency, SLL, and effective aperture (total area of radiating elements) for different aperture sizes (values of $M$). Fig. \ref{Analysis1}(b) shows that while the half beamwidth decreases for larger and larger apertures, the beam efficiency remains relatively constant. Furthermore, Fig. \ref{Analysis1}(d) shows that SLL decreases as the aperture size gets larger and the effective aperture size also increases. Thus, multi-annular-ring apertures are good candidates for sparse phased-array receiver applications where collection area and SLL are important performance metrics. Moreover, these plots suggest that for a given $(r_1,N_1)$ pair, the beam performance improves while the beam efficiency remains constant. This provides a design methodology for multi-annular-ring apertures. We can proceed and calculate the largest apertures that can be constructed using these linear density annular-ring apertures. Assuming the waveguide and radiator size and pitch limitations of silicon photonics platforms mentioned above, we can plot the minimum achievable beamwidth for a given beam efficiency for single-layer and two-layer photonic platforms, as shown in Fig. \ref{Analysis1}(c). Thus, a multi-annular-ring OPA implemented in a single layer photonic platform with $1\%$ beam efficiency can resolve 9,300 points over the entire FOV. A two-layer photonic platform with the same efficiency approximately doubles the number of resolvable points to 18,500 in the far-field.

\section{Multi-Annular-Ring OPA Implementation}      

\begin{figure}[!b]
	\includegraphics[width=1\linewidth]{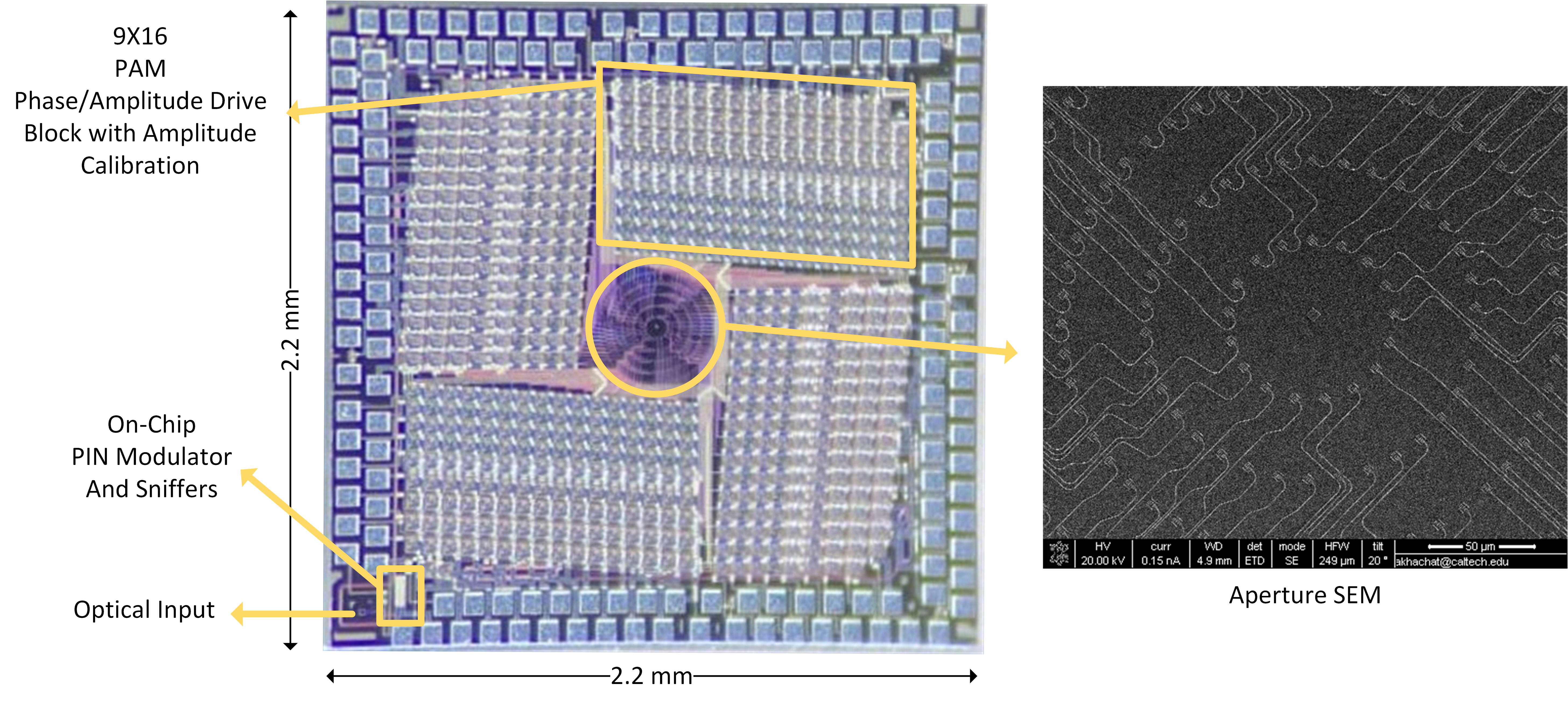}
	\caption{Multi-annular-ring aperture OPA system. Die photo of the proposed design and SEM photo of the aperture. Phase and amplitude modulators are grouped into four blocks for symmetric layout.}
	\label{COPA_layout}
\end{figure}

To demonstrate the beamforming and beam steering capability of multi-annular-ring aperture OPAs, we implemented a five-annular-ring aperture OPA system with active beamforming in Advanced Micro Foundry's (AMF) standard photonics process. The die photo of this system is shown in Fig. \ref{COPA_layout}. This OPA  has a \SI{400}{\micro \meter} diameter annular-ring aperture with 255 radiating elements with complete phase and amplitude control. The thermo-optic phase and amplitude modulators are distributed into four blocks and electrically connected in a row-column fashion to reduce the electrical interconnect density from order $N^2$ to order $N$. Amplitude modulation is achieved via a $1:2^N$ tunable amplitude distribution network which can be calibrated with only $N+1$ integrated sniffer photodiodes. The design details of these three sub-blocks are examined in the following sub-sections.

\subsection{Implemented Aperture}
\begin{figure}[!t]
	\includegraphics[width=1\linewidth]{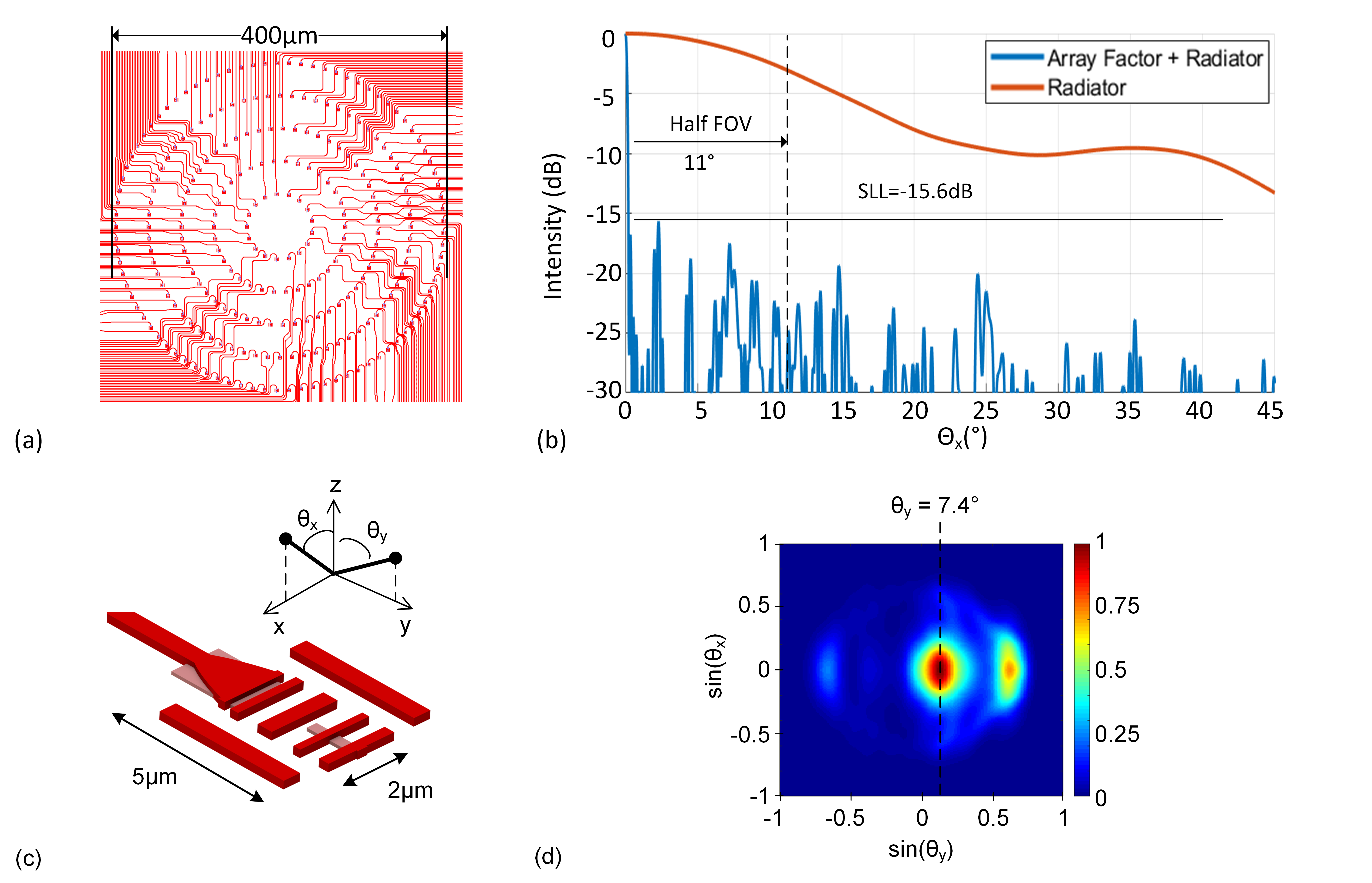}
	\caption{ (a) The layout and signal distribution for a 255-element annular-ring aperture. (b) Full AF of the aperture for $\phi=0$ including the radiator's beam pattern. (c) The custom radiating element used in this OPA. (d) The far-field radiation pattern of the radiating element.}
	\label{AF_COPA}
\end{figure}

This multi-annular-ring OPA array factor was optimized based on a \SI{1}{\micro \meter} minimum pitch in the waveguide routing with $\SI{2}{\micro \meter}\times \SI{5}{\micro \meter}$ compact photonics radiators (Fig. \ref{AF_COPA}(c)). Based on the analysis in section \ref{LinearDensityTheory}, we incorporated a five-annular-ring ($M=5$) aperture with a linear density ($r_1=\SI{40}{\micro \meter}$ and $N_1=17$). As a result,  $17,34,51,68$, and $85$ radiators are placed on rings with respective radii of $\SI{40}{\micro \meter}, \SI{80}{\micro \meter}, \SI{120}{\micro \meter}, \SI{160}{\micro \meter}$, and $\SI{200}{\micro \meter}$, as shown in Fig. \ref{AF_COPA}(a). This multi-annular-ring aperture maintains the Bessel form for a $2^\circ$ FOV with $0.2^\circ$ theoretical beamwidth with no grating lobes in the full FOV. The cross-section of the radiation pattern for this OPA is shown in Fig. \ref{AF_COPA}(b). 

\par
The radiating element used in this aperture has a \SI{3}{\decibel} far-field beamwidth of $23^\circ \times 16.3^\circ$ with over $50\%$ peak radiation efficiency at the optimum angle of $7.4^\circ$ \cite{Reza:JSSC}. Furthermore, the \SI{1}{\decibel} spectral bandwidth of the radiator is more than \SI{400}{nm}, which makes these radiators insensitive to temperature and wavelength variations. As a result, the FOV of this aperture is limited by the FOV of the radiating element with a maximum SLL of \SI{-15.6}{\decibel} as shown in Fig. \ref{AF_COPA}(b). Furthermore, the theoretical resolution of this OPA is $115\times81=9315$ total points.

\subsection{Row-Column Drive}

\begin{figure}[!t]
	\includegraphics[width=1\linewidth]{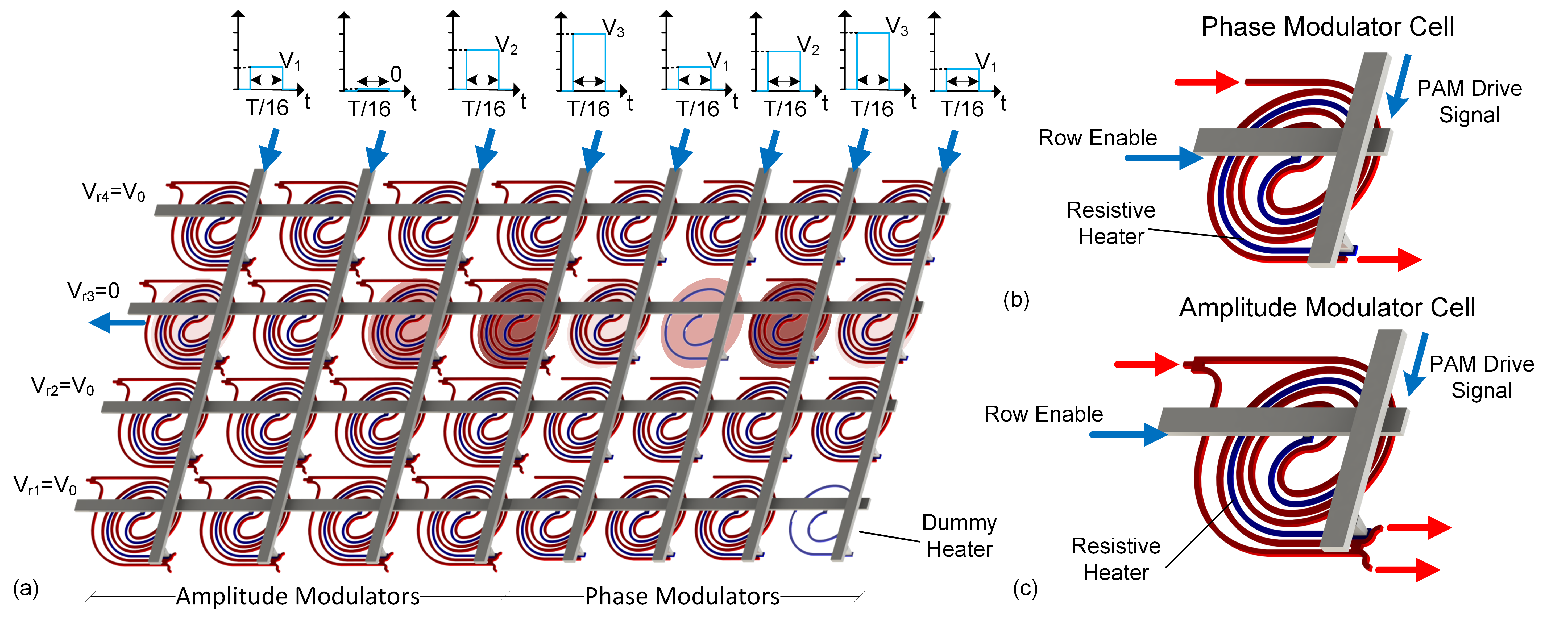}
	\caption{(a) Row-column drive scheme for amplitude and phase modulators. Each row is active for $T/16$ of the programming cycle and programs all elements in the row. (b) Implemented phase modulator using a compact spiral thermo-optic modulator. (c) Implemented amplitude modulator based on a tunable MZI splitter using a compact spiral modulator.}
	\label{COPA_PAMdrive}
\end{figure}

The amplitude and phase modulators for the 255-element array are divided into four blocks. Both amplitude and phase modulators incorporate a compact spiral thermo-optic phase shifter design to reduce the device footprint and increase isolation between radiating elements \cite{Reza:JSSC} as shown in Fig. \ref{COPA_PAMdrive}(b)-(c). The amplitude modulation is achieved by a cascade of tunable optical couplers that split the light into 64 branches. This tunable power splitter requires $63$ tunable couplers. The output of each tunable coupler passes through phase shifters before reaching the radiating elements to adjust the relative phase between elements. These $127$ active components are electrically connected in a row-column fashion, as shown in Fig. \ref{COPA_PAMdrive}(a). The tunable couplers are arranged in a $16 \times 4$ grid. There are $64$ phase shifters in each block. In addition, there are $16$ distributed dummy thermal phase shifters placed among the phase shifters to reduce the thermal gradient in the substrate. These phase shifters and dummy heaters are arranged in a $16 \times 5$ grid resulting in  $4(16+9)=100$ drive nodes for the full OPA.
\par
To achieve independent control of all phase and amplitude modulators, each of these four blocks is programmed by a time-domain demultiplexing technique that utilizes the thermal memory of these phase shifters. In this scheme, the modulators are programmed in continuous cycles. Each programming cycle, $T$, is divided into $16$ time slots. During each time slot ($T/16$), only one row of the phase shifter matrix is programmed using pulse-amplitude-modulation (PAM) by delivery of the desired thermal power to elements in the active row and keeping the remaining rows of the modulators in the off state. This is achieved by including a series diode with the resistive heater to prevent the flow of current through the turned-off columns of the array by keeping the diode in reverse bias. Since these thermal phase shifters have a bandwidth on the order of several kHz, having a programming period $T$ in the MHz range ensures that all phase shifters maintain their temperature or programmed phase value. This row-column drive methodology reduces the required number of electrical interconnect nodes for an $N\times M=9\times16$ array from $144$ nodes to $N+M=25$ nodes.

\subsection{Amplitude Modulation with Simplified On-Chip Calibration}

\begin{figure}[!t]
	\includegraphics[width=1\linewidth]{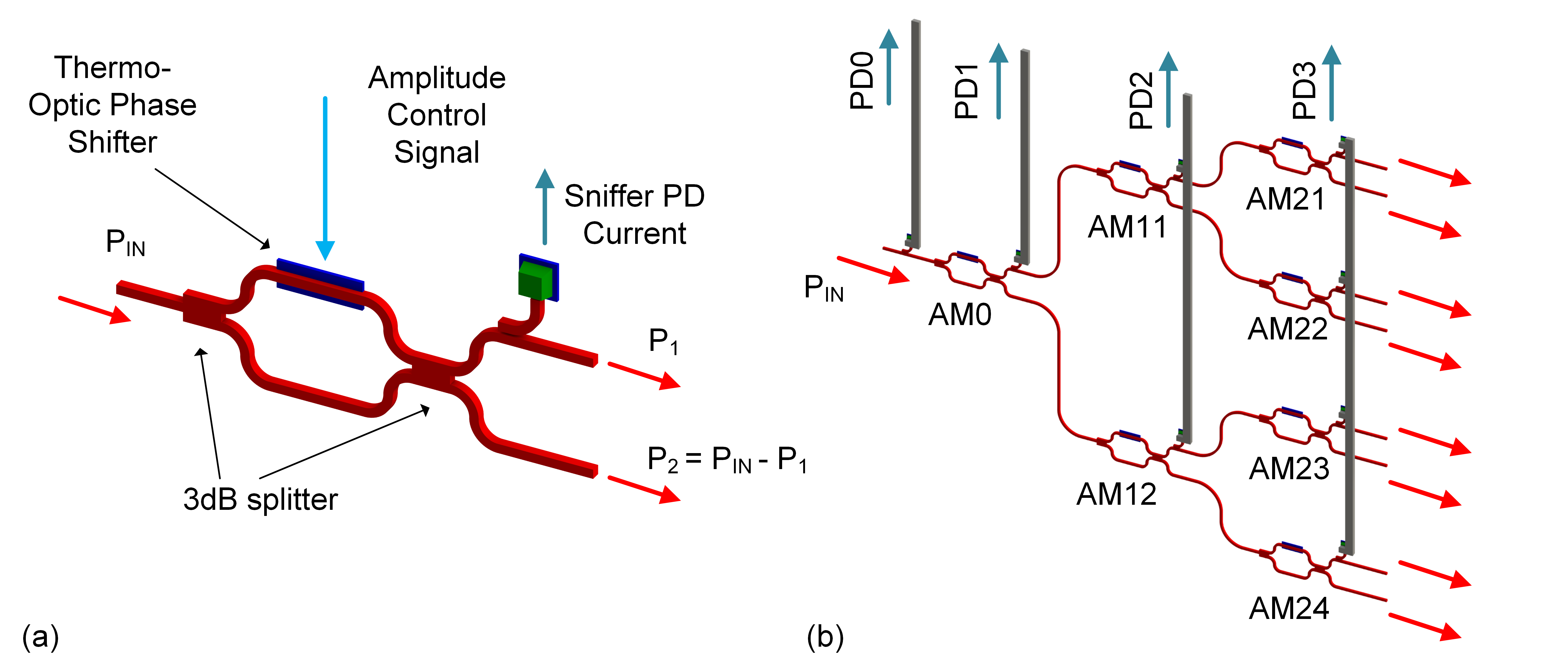}
	\caption{Tunable amplitude modulation with calibration feedback. (a) Unit tunable power splitter with $1\%$ sniffer output for control. (b) $1:8$ tunable power splitter with reduced sensing interconnect complexity.}
	\label{COPA_AMc}
\end{figure}

Amplitude modulation for the 255 elements in this design is achieved by cascading $8$ one to two tunable optical couplers (Fig. \ref{COPA_AMc}(a)). This design is advantageous over the conventional approach of dedicating one amplitude attenuator per signal path \cite{Abediasl:15} since for different amplitude configurations, the power is redistributed between different paths, and the total power delivered to the aperture remains constant. This distribution method can be used to deliver equal power to all elements or to achieve amplitude apodization for reduced SLL. Each output of these amplitude modulators has a $1\%$ power splitter and a compact sniffer photodetector for on-chip calibration. The output of these photodiodes can be used to perform a one-time calibration to correct for fabrication mismatches in the tunable amplitude coupler and determine the drive voltage requirement for the amplitude modulators. 
\par
It can be noted that this calibration requires $2^N-1$ sniffer diodes and $2^N-1$ sensing nodes for a $1:2^N$ splitter array. However, since this calibration needs to be performed only once, it is possible to reduce the number of required sensing nodes to $N+1$ by combing the output current of detectors in each stage and using a sequential calibration methodology. This methodology is described using the $1:8$ splitter in Fig. \ref{COPA_AMc}(b). The output of the first sensing stage (PD1) can be used to calibrate the tunable coupler (AM0) in the first splitting stage. Afterward, AM0 is configured in order that all power is delivered to AM11. As a result, the combined output of the second sensing stage (PD2) corresponds only to the top branch (AM11). This sensing output can be used to calibrate AM11. Iterating this process for all the remaining amplitude modulators allows full calibration of this $1:8$ splitter using only $4$ sensing nodes. For the 255-element array, only $9$ sensing nodes are required, which significantly reduces the sensing interconnect complexity.

\par
In this OPA system implementation, the power is coupled into the aperture using a lensed grating coupler. An integrated PIN modulator isolates the coupled light from the scattered reflected light from the substrate. Integrated PTAT sensors can be used to measure the substrate temperature gradient across the chip. The optical path-length mismatch between radiating elements is compensated for by incorporating additional delay lines in the chip. The entire system has a $\SI{2.2}{\milli \meter}\times \SI{2.2}{\milli \meter}$ area.

\section{Measurement Result} 

The amplitude and phase modulator unit cell with the compact spiral phase shifters was characterized in \cite{Reza:JSSC}. These modulators have \SI{19}{\kilo \hertz} electro-optical bandwidth and require \SI{21.2}{\milli \watt} for a $2\pi$ phase shift. For the beamforming optimization of the proposed annular-ring-aperture OPA, $64$ PAM drivers with 10 bits of resolution and $36$ row-enable switches with \SI{10}{\volt} maximum swing and 10ns rise and fall time were used. The row-column drive circuits operates at a \SI{1.56}{\mega \hertz} repetition rate which is sufficiently larger that than the thermal time-constant of the spiral modulators. 

\begin{figure}[!t]
	\includegraphics[width=1\linewidth]{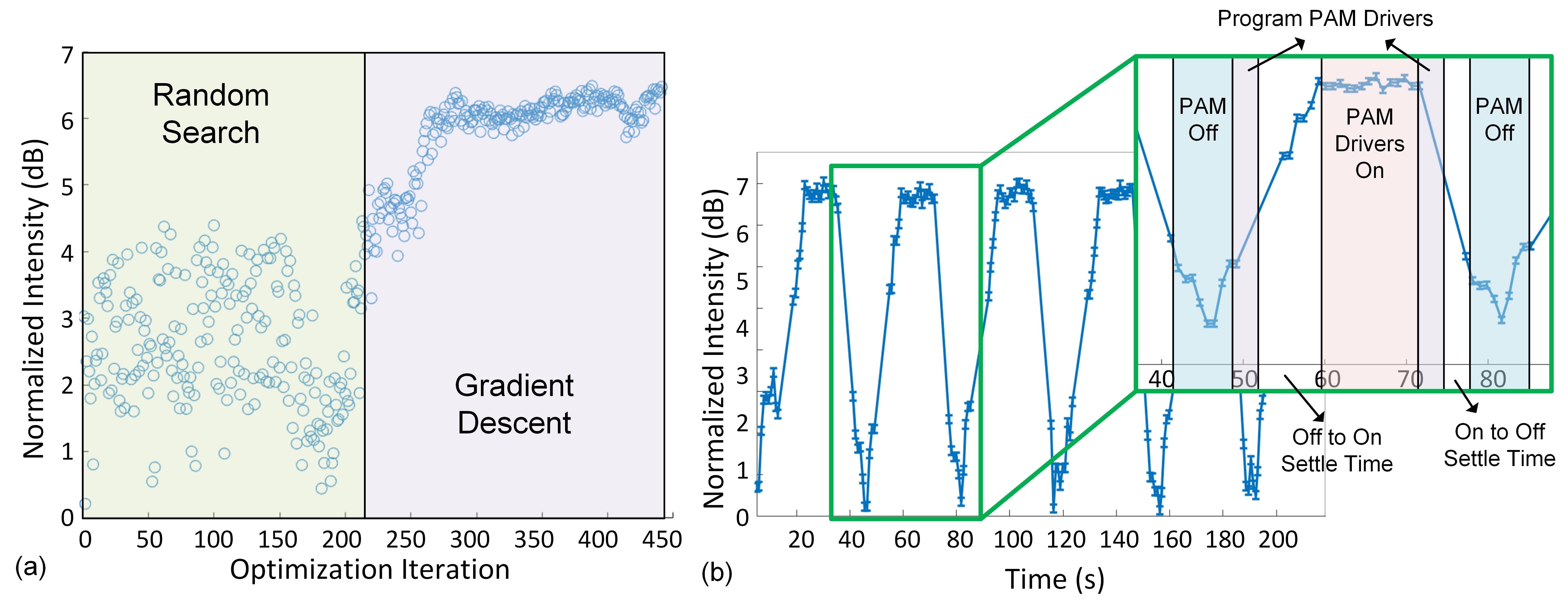}
	\caption{(a) Beam power optimization progression after 450 iterations. (b) Dynamic stability and repeatability of the annular-ring OPA. The PAM drivers can maintain the optimized setting with less than \SI{0.4}{\decibel} variations.}
	\label{AnnularRing_OPT}
\end{figure}

\par
The far-field radiation pattern of the OPA was captured using a custom optical far-field radiation measurement setup. This apparatus moves a compact InGaAs photodetector along the arc at a fixed distance with respect to the OPA chip and captures the far-field pattern point by point. The \SI{1.55}{\micro \meter} light coupled into the chip is modulated at \SI{1.1}{\mega \hertz} by the integrated on-chip PIN modulator to isolate the radiated power from the stray uncoupled light. The far-field radiation power was filtered and quantized using a spectrum analyzer with \SI{10}{\hertz} resolution bandwidth. The collected power was normalized with respect to the power coupled into the chip to remove any changes in power due to the slight variations in the position of the input fiber during the measurement.

\begin{figure}[!b]
	\includegraphics[width=1\linewidth]{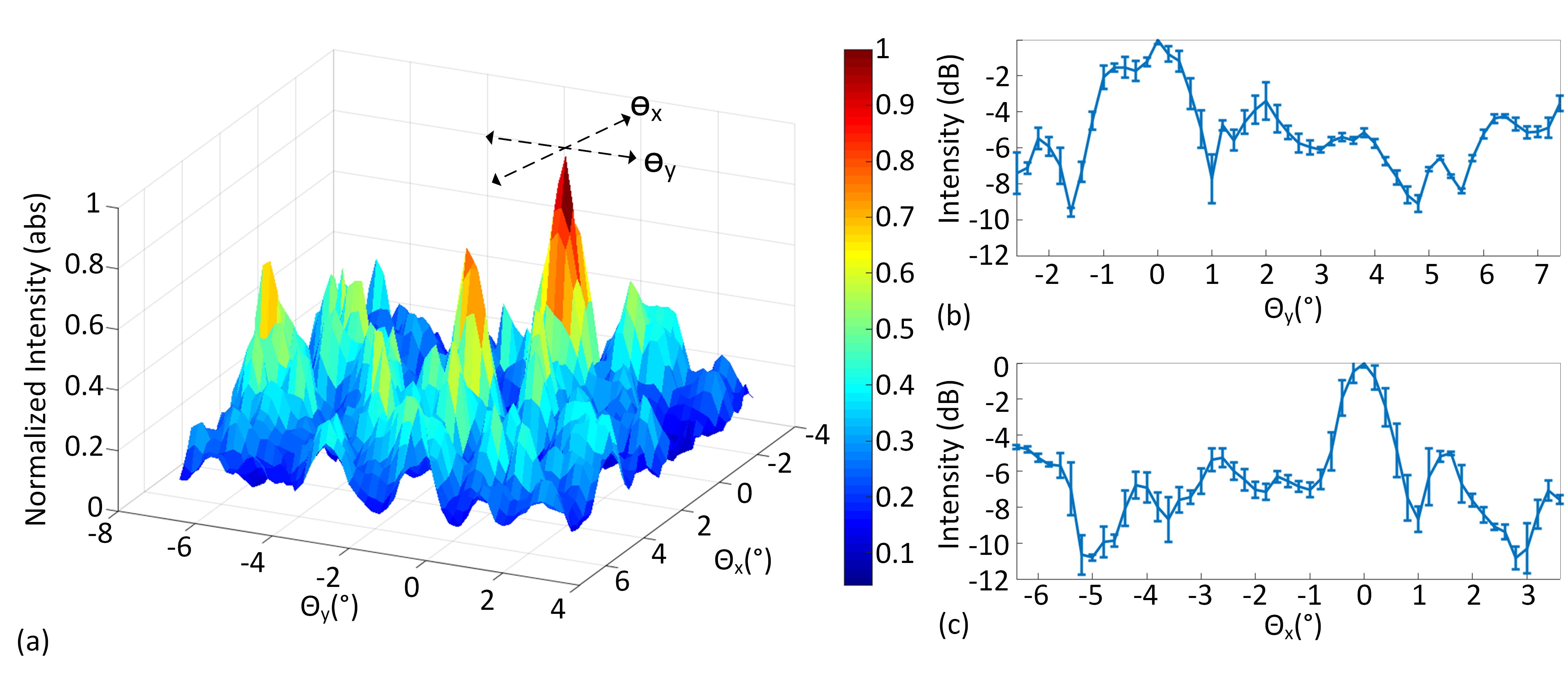}
	\caption{2D beamforming demonstration. (a) Two-dimensional beam pattern was measured for the optimized direction $(0,0)$. (b) 1D cross-sections of the beam pattern in $\theta_x$ ($\phi=0$ plane). (c) 1D cross-sections of the beam pattern in $\theta_y$ ($\phi=90^\circ$ plane).}
	\label{PolarOPA_Meas2D}
\end{figure}
\par
All of the phase and amplitude modulators in the array were optimized concurrently using the $64$ PAM drivers. These PAM drivers switch between $16$ different values with 10-bit resolution for the $16$ different columns. The large number of driver settings creates a very large search space for finding the optimized beam. In order to address this problem, we first performed a randomized optimization and recorded the broadside beam power value. After $200$ iterations, the randomized search found a PAM setting with \SI{4}{\decibel} increased signal power in the main beam. Afterward, a modified gradient search continues to optimize the main beam power from the best random-optimized beam. After $250$ iterations, gradient descent improves the beam by another \SI{2}{\decibel}. The optimization progression is shown in Fig. \ref{AnnularRing_OPT}(a).

\begin{figure}[!t]
	\includegraphics[width=1\linewidth]{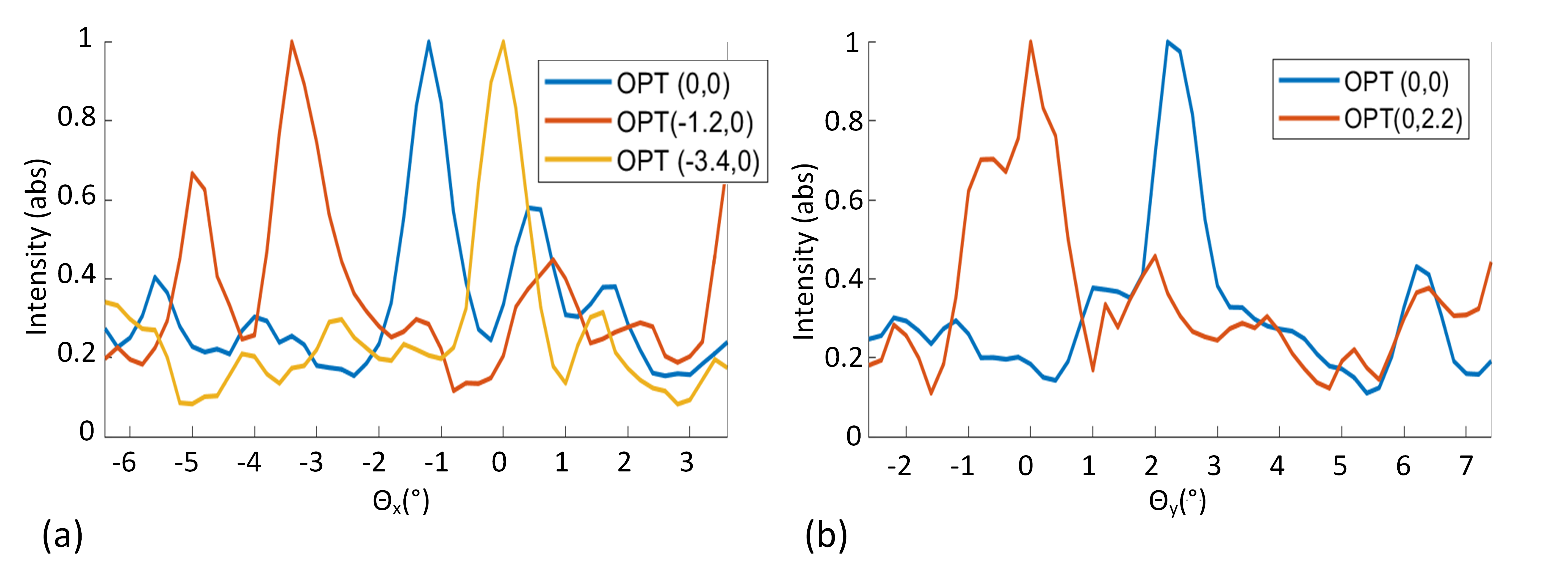}
	\caption{Cross-sectional view of the beam pattern for several directions. (a) Phase-shifter enabled beam steering in $\theta_x$ ($\phi=0$ plane) for two additional directions. (b) Phase-shifter enabled beam steering in $\theta_y$ ($\phi=90^\circ$ plane) for an additional direction.}
	\label{PolarOPA_Meas1D}
\end{figure}
\par
To verify the dynamic beamforming stability and repeatability of this OPA, we switched between the setting with all PAM drivers off and the optimum PAM setting for the broadside and recorded the dynamic changes in the normalized main beam power (with respect to the power coupled into the chip) as a function of time. The dynamic stability of the PAM drive is shown in Fig. \ref{AnnularRing_OPT}(b). It takes $2$ seconds to program the PAM drivers. The thermal gradient variations on the chip settle in approximating $8$ seconds. The PAM drivers are able to consistently program the phase and amplitude modulators through the row-column drive scheme with less than \SI{0.4}{\decibel} ripple in the peak power.

\par
Afterward, we measured the optimized beam pattern using the aforementioned optical far-field measurement setup. The InGaAs photodector was rotated along the arc in $(\theta_x,\theta_y)$ directions at a 5cm distance. A 2D scan of the optimized beam pattern for the broadside direction $(\theta_x,\theta_y)=(0,0)$ in Fig. \ref{PolarOPA_Meas2D}(a) shows no grating lobes in the measurement FOV of $(10^\circ,10^\circ)$ limited by the measurement setup. This measurement result was repeated three times and the cross-sectional view of the normalized broadside beam pattern with standard deviation for $\theta_y=0$ and $\theta_x=0$ are shown in Fig. \ref{PolarOPA_Meas2D}(b) and (c) respectively. The optimization method used here was able to reduce the SLL to \SI{-4}{\decibel}. Finally, we demonstrated the beam steering capability of this annular-ring aperture by optimizing the beam in several other directions. The InGaAs detector was positioned at $(\theta_x,\theta_y)=(-1.2,0)$  and $(\theta_x,\theta_y)=(-3.4,0)$ to demonstrate beam steering in $\theta_x$ direction and it was also positioned at $(\theta_x,\theta_y)=(0,2.2)$ to demonstrate beam steering in $\theta_y$ direction. The cross-sectional views of these steered beams are shown in Fig. \ref{PolarOPA_Meas1D}.

\section{Conclusion}
In this work, we analyzed the advantages and design trade-offs of fixed linear density multi-annular-ring-aperture OPAs. Such OPAs reduce the sparse array design complexity due to their symmetric nature and permit signal distribution in a planar photonics process. Furthermore, the linear density of the elements in the OPA maintains a constant power efficiency for larger apertures reducing the beamwidth and the SLL. In a standard photonics process, we implemented an annular-ring aperture with 255 radiating elements and 510 phase and amplitude modulators. The proposed design uses a row-column drive, which reduces the electrical interconnect complexity from $N^2$ to $2N$. Furthermore, we measured the row-column drive programming reliability to be better than \SI{0.4}{\decibel}. Finally, we measured the far-field radiation pattern of the aperture for several directions, demonstrating the beamforming and beam steering capability for this aperture.
\par

\section*{Acknowledgments}
The authors would like to acknowledge Behrooz Abiri and Parham Porsandeh Khial for their valuable inputs in the design and analysis of this work.

\section*{Disclosures}

The authors disclose no conflict of interest.

\bibliography{PolarOPA}


\end{document}